\documentclass[aip,reprint,showpacs,showkeys]{revtex4-1}
\usepackage[T1]{fontenc}
\usepackage{times}
\usepackage{graphicx,color}
\usepackage{amsfonts,amsmath,amssymb,amsbsy}
\usepackage[colorlinks=true, linkcolor=blue, citecolor=blue, urlcolor=blue]{hyperref}

\let\mathbf=\boldsymbol

\def\emph#1{\textcolor{blue}{#1}}
\begin{document}

\title{Dynamics of an elliptical ferromagnetic skyrmion driven by the spin-orbit torque}

\author{Jing Xia}
\thanks{These authors contributed equally to this work.}
\affiliation{School of Science and Engineering, The Chinese University of Hong Kong, Shenzhen, Guangdong 518172, China}

\author{Xichao Zhang}
\thanks{These authors contributed equally to this work.}
\affiliation{School of Science and Engineering, The Chinese University of Hong Kong, Shenzhen, Guangdong 518172, China}

\author{Motohiko Ezawa}
\affiliation{Department of Applied Physics, The University of Tokyo, 7-3-1 Hongo, Tokyo 113-8656, Japan}

\author{Qiming Shao}
\affiliation{Department of ECE, The Hong Kong University of Science and Technology, Clear Water Bay, Kowloon, Hong Kong, China}

\author{Xiaoxi Liu}
\affiliation{Department of Electrical and Computer Engineering, Shinshu University, 4-17-1 Wakasato, Nagano 380-8553, Japan}

\author{Yan Zhou}
\email[E-mail:~]{zhouyan@cuhk.edu.cn}
\affiliation{School of Science and Engineering, The Chinese University of Hong Kong, Shenzhen, Guangdong 518172, China}

\begin{abstract}
Magnetic skyrmion is a promising building block for developing information storage and computing devices. It can be stabilized in a ferromagnetic thin film with the Dzyaloshinskii-Moriya interaction (DMI). The moving ferromagnetic skyrmion may show the skyrmion Hall effect, that is, the skyrmion shows a transverse shift when it is driven by a spin current. Here, we numerically and theoretically study the current-driven dynamics of a ferromagnetic nanoscale skyrmion in the presence of the anisotropic DMI, where the skyrmion has an elliptical shape. The skyrmion Hall effect of the elliptical skyrmion is investigated. It is found that the skyrmion Hall angle can be controlled by tuning the profile of elliptical skyrmion.
Our results reveal the relation between the skyrmion shape and the skyrmion Hall effect, which could be useful for building skyrmion-based spintronic devices with preferred skyrmion Hall angle.
Also, our results provide a method for the minimization of skyrmion Hall angle for applications based on in-line motion of skyrmions.
\end{abstract}

\date{6 January 2020}

\preprint{\textsl{To be submitted to Appl. Phys. Lett.}}

\maketitle


Magnetic skyrmions are topologically non-trivial spin textures,~\cite{Roszler_NATURE2006,Nagaosa_NNANO2013,Finocchio_JPD2016,Kang_PIEEE2016,Wiesendanger_Review2016,Fert_NATREVMAT2017,Wanjun_PHYSREP2017,Everschor_JAP2018,Xichao_ARXIV2019} which can be used to build future memories,~\cite{Sampaio_NNANO2013,Tomasello_SREP2014,Guoqiang_NL2017,Muller_NJP2017} logic computing devices,~\cite{Xichao_SREP2015B} and bio-inspired computing devices.~\cite{Yangqi_NANO2017,Lisai_NANO2017,Prychynenko_PRAPPL2018,WOO_arXiv2019}
The magnetic skyrmion in a ferromagnetic thin film can be created and driven into motion by spin currents.~\cite{Yin_PRB2016} However, it may experience the skyrmion Hall effect,~\cite{Zang_PRL2011,Wanjun_NPHYS2017,Litzius_NPHYS2017} that is, the skyrmion shows a transverse displacement due to the topological Magnus force acted on the skyrmion.
In order to build some skyrmion-based spintronic devices using the in-line motion feature of skyrmions, it is necessary to eliminate the skyrmion Hall effect since the skyrmion Hall effect may lead to the destruction of skyrmions at sample edges.
Several proposals have been proposed to eliminate the skyrmion Hall effect, for examples, the skyrmion Hall effect can be avoided in the synthetic antiferromagnetic bilayers~\cite{Xichao_NCOMMS2016,Xichao_PRB2016B} and antiferromagnetic thin films.~\cite{Barker_PRL2016,Zhang_SREP2016}

On the other hand, the Dzyaloshinskii-Moriya interaction (DMI) is an essential interaction to stabilize the magnetic skyrmion in bulk and thin-film materials.~\cite{Bogdanov_JMMM1994,Nagaosa_NNANO2013,Finocchio_JPD2016,Wiesendanger_Review2016,Fert_NATREVMAT2017,Wanjun_PHYSREP2017,Everschor_JAP2018,Xichao_ARXIV2019}
The interfical DMI~\cite{Romming_SCIENCE2013,Wanjun_SCIENCE2015,Woo_NMATER2016,MoreauLuchaire_NNANO2016,Boulle_NNANO2016,Nozaki_APL2019} can be induced at the interface between a heavy metal and ferromagnet. The bulk DMI~\cite{Muhlbauer_SCIENCE2009,Yu_NATURE2010,Du_NCOMMS2015} can be induced by introducing impurities with large spin-orbit coupling in ferromagnets. Both the two types of DMIs are arisen by the inversion-symmetry-broken structure.

Recently, elliptical skyrmions have been found in some experiments.~\cite{Hsu_NNANO2017,Hagemeister_PRB2016,Nagase_PRL2019}
It is found that the DMI can be anisotropic in the Co/W(110) stack with a $C_{2v}$ symmetry, where the DMI strength is $2-3$ times larger along bcc[$\bar{\text{1}}$10] than along bcc[001].~\cite{Camosi_PRB2017}
When the strength of the DMI in two directions are different, the shape of skyrmion is elliptical rather than circular.~\cite{Gungordu_PRB2016,Osorio_PRB2019}
Recent studies show that the shape of skyrmion has an impact on spin wave modes and skyrmion Hall effect.~\cite{Liu_JMMM2018,Juge_PRAPPLIED2019}
In this work, we report the current-driven dynamics of an elliptical skyrmion, which is stabilized by the anisotropic DMI in a ferromagnetic thin film. The motion of the elliptical skyrmion driven by the spin-orbit torque is investigated by both numerical and theoretical methods. It is found that the skyrmion Hall effect of the elliptical skyrmion can be reduced to some extent compared to the case of circular skyrmion.


We perform micromagnetic simulations by using the Object Oriented MicroMagnetic Framework (OOMMF) developed at the National Institute of Standards and Technology (NIST).~\cite{OOMMF} In the presence of the spin-orbit torque, the magnetization dynamics is governed by the Landau-Lifshitz-Gilbert (LLG) equation augmented with a damping-like torque~\cite{OOMMF,Tomasello_SREP2014}
\begin{equation}
\begin{split}
\label{eq:LLGS-CPP}
\frac{d\boldsymbol{M}}{dt}=&-\gamma_{0}\boldsymbol{M}\times\boldsymbol{H}_{\text{eff}}+\frac{\alpha}{M_{\text{S}}}(\boldsymbol{M}\times\frac{d\boldsymbol{M}}{dt}) \\
&+\frac{u}{aM_{\text{S}}}(\boldsymbol{M}\times \boldsymbol{p}\times \boldsymbol{M}),
\end{split}
\end{equation}
where $\boldsymbol{M}$ is the magnetization, $M_{\text{S}}=|\boldsymbol{M}|$ is the saturation magnetization, $t$ is the time, $\gamma_{\text{0}}$ is the gyromagnetic ratio with absolute value, and $\alpha$ is the Gilbert damping coefficient. $\boldsymbol{H}_{\text{eff}}$ is the effective field, which reads $\boldsymbol{H}_{\text{eff}}=-\mu_{0}^{-1}\frac{\partial E}{\partial \boldsymbol{M}}$.
The average energy density $E$ contains the Heisenberg exchange, the perpendicular magnetic anisotropy (PMA), the demagnetization, and the DMI energy terms. For the anisotropic DMI, the DMI energy can be expressed as~\cite{HuangSiying_PRB2017}
\begin{equation}
\begin{split}
\label{eq:E} 
E_\text{DM}=&\frac{D_x}{M_{\text{S}}^{2}}(M_{z}\frac{\partial M_{x}}{\partial x}-M_{x}\frac{\partial M_{z}}{\partial x}) \\ 
&+\frac{D_y}{M_{\text{S}}^{2}}(M_{z}\frac{\partial M_{y}}{\partial y}-M_{y}\frac{\partial M_{z}}{\partial y}), 
\end{split}
\end{equation}
where $D_x$ and $D_y$ are DMI energy constants. The $M_x$, $M_y$ and $M_z$ are the three Cartesian components of the magnetization $\boldsymbol{M}$.
$u=|\frac{\gamma_{0}\hbar}{\mu_{0}e}|\frac{j\theta_{\text{SH}}}{2M_{\text{S}}}$ is the spin torque coefficient, and $\boldsymbol{p}$ stands for the unit spin polarization direction. $\hbar$ is the reduced Planck constant, $e$ is the electron charge, $j$ is the applied current density, and $\theta_{\text{SH}}$ is the spin Hall angle.

In our simulations, we model an ultra-thin ferromagnetic film with a side length of $200$ nm and a thickness of $a=0.4$ nm. The mesh size is set as $1 \times 1 \times 0.4$ nm$^3$. The intrinsic magnetic material parameters are adopted from Ref.~\onlinecite{Sampaio_NNANO2013}: the ferromagnetic exchange constant $A=15$ pJ/m, saturation magnetization $M_{\text{S}}=0.58$ MA/m, and PMA constant $K=0.8$ MJ/m$^3$. $D_x$ and $D_y$ vary from $2.5$ mJ/m$^2$ to $3.7$ mJ/m$^2$.
For the motion of skyrmion, the driving current density is set as $15 \times 10^{10}$ A/m$^2$. We also assume that $\boldsymbol{p}=+\hat{y}$ and $\theta_{\text{SH}}=0.08$. The injection duration of driving current is fixed at $7$ ns. 

\begin{figure}[t]
\centerline{\includegraphics[width=0.4\textwidth]{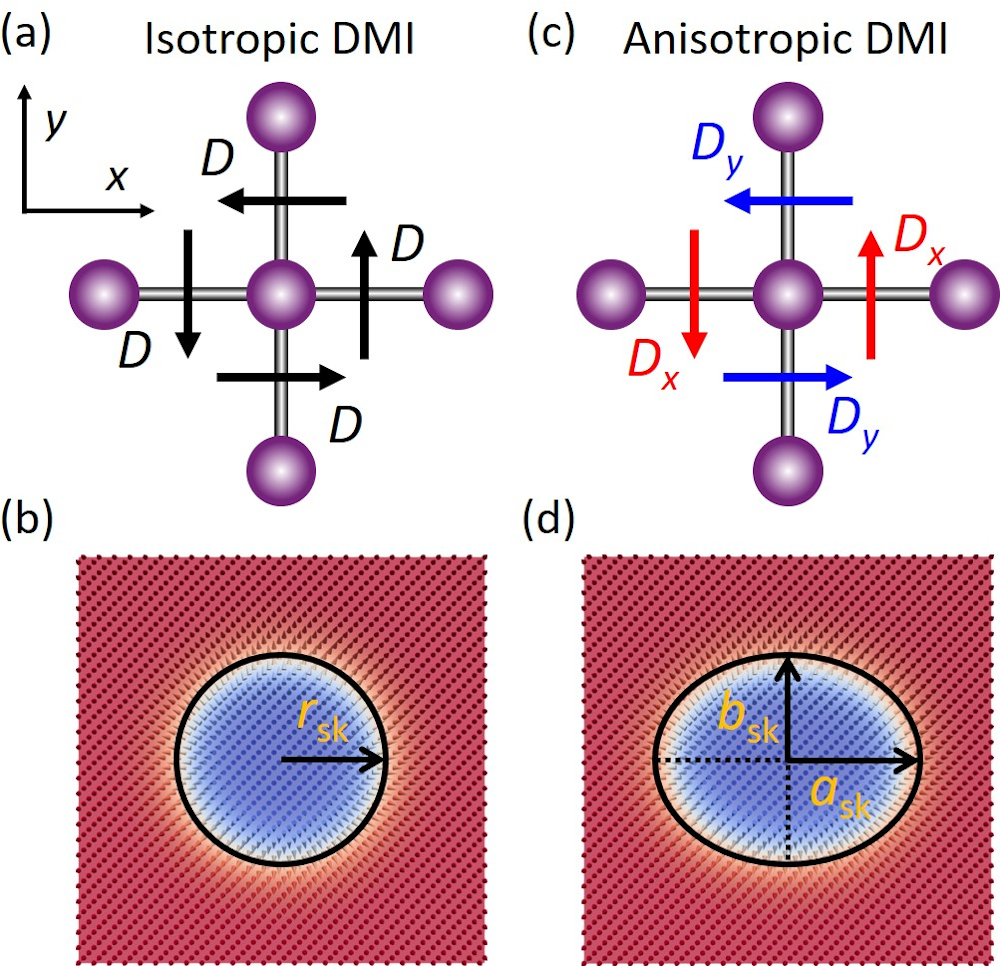}}
\caption{
(a) Schematic of the isotropic DMI. $D_x$ and $D_y$ represent the coefficients of the DMI in the $x$-axis and $y$-axis, respectively. For the isotropic case, $D_x=D_y=D$.
(b) The circular skyrmion stabilized by the isotropic DMI. The out-of-plane magnetization component is represented by the red ($+z$)-white ($0$)-blue ($-z$) color scale.
(c) Schematic of the anisotropic DMI, where $D_x\neq D_y$.
(d) The elliptical skyrmion stabilized by the anisotropic DMI with $D_x < D_y$.
The black circle and oval in (b) and (d) are the contours of skyrmions ($m_z=0$). $r_\text{sk}$ denotes the radius of the circular skyrmion. $a_\text{sk}$ and $b_\text{sk}$ describe the size and shape of elliptical skyrmion.
}
\label{FIG1}
\end{figure}

Figure~\ref{FIG1} illustrates the isotropic and anisotropy DMIs and the corresponding skyrmion configurations.
For the isotropic case, $D_x=D_y=D$ and a circular skyrmion will be obtained for relaxed system, as shown in Fig.~\ref{FIG1}(b).
For the anisotropic case, $D_x \neq D_y$. The relaxed skyrmion will be deformed to have an elliptical shape, as shown in Fig.~\ref{FIG1}(d).

\begin{figure}[t]
\centerline{\includegraphics[width=0.50\textwidth]{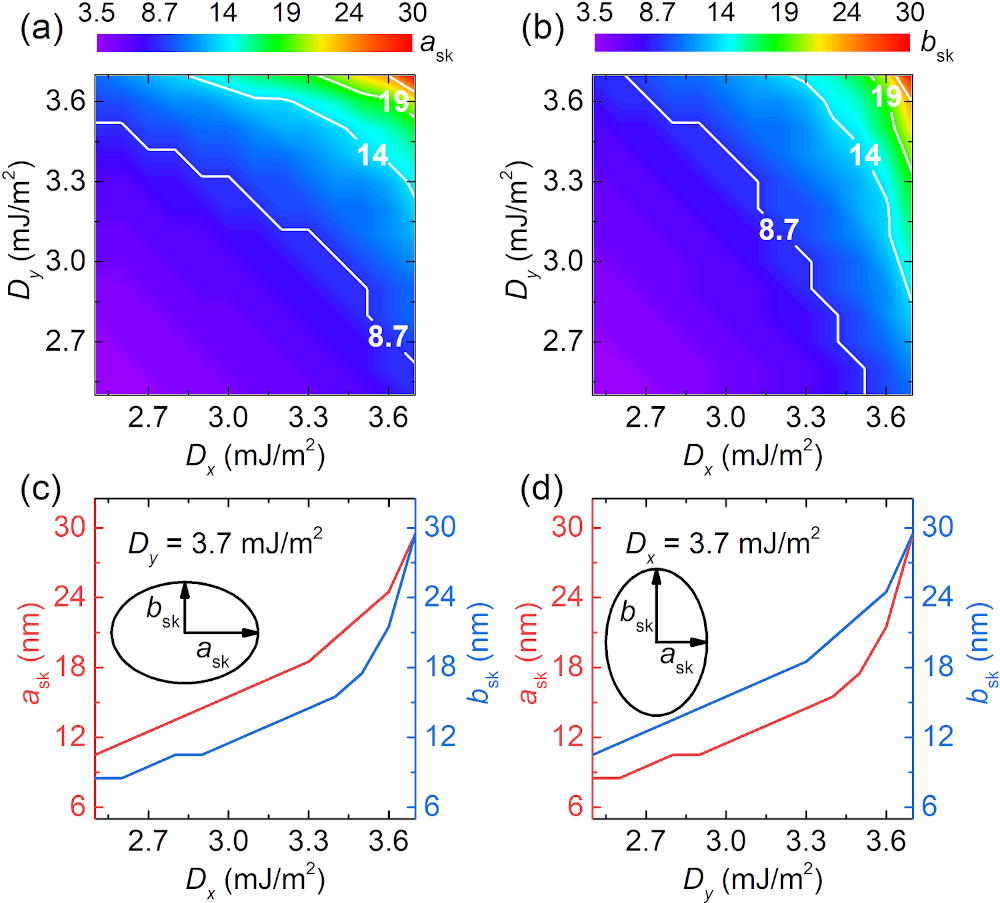}}
\caption{
(a) $a_\text{sk}$ as functions of $D_x$ and $D_y$.
(b) $b_\text{sk}$ as functions of $D_x$ and $D_y$.
(c) $a_\text{sk}$ and $b_\text{sk}$ as function of $D_x$ when $D_y = 3.7$ mJ/m$^2$.
(d) $a_\text{sk}$ and $b_\text{sk}$ as function of $D_y$ when $D_x = 3.7$ mJ/m$^2$.
}
\label{FIG2}
\end{figure}


We first micromagnetically simulate the relaxed configuration of skyrmion in the sample of $200~\text{nm} \times 200~\text{nm}$.
Figure~\ref{FIG2}(a) and \ref{FIG2}(b) show that both $a_\text{sk}$ and $b_\text{sk}$ increases with $D_x$ and $D_y$ (see Supplementary Information).
Moreover, it is found that the anisotropic DMI leads to the formation of an elliptical skyrmion. For example, as shown in Fig.~\ref{FIG2}(c), $a_\text{sk}$ is larger than $b_\text{sk}$ when $D_x<D_y$ while smaller when $D_x>D_y$ [see Fig.~\ref{FIG2}(d)], which are consistent with the results in Ref.~\onlinecite{Osorio_PRB2019} (see Supplementary Information for the relation between $a_\text{sk}/b_\text{sk}$ and $D_y/D_x$).
When $D_x=D_y=3.7$ mJ/m$^2$, a circular skyrmion is obtained with a radius of $29.5$ nm.
The radius of circular skyrmion increases from $3.5$ nm to $29.5$ nm when the strength of DMI varies from $2.5$ mJ/m$^2$ to $3.7$ mJ/m$^2$. This results agree well with the dependence of skyrmion radius on $D$, $r_\text{sk}=\pi D \sqrt{A/(16AK_\text{eff}^2-\pi^2D^2K_\text{eff})}$ with $K_\text{eff}=K-\mu_\text{0}M_\text{S}^2/2$.~\cite{WANGXS2018} It should be mentioned that the skyrmion number $Q$ for the elliptical and circular skyrmions are the same.

\begin{figure}[t]
\centerline{\includegraphics[width=0.45\textwidth]{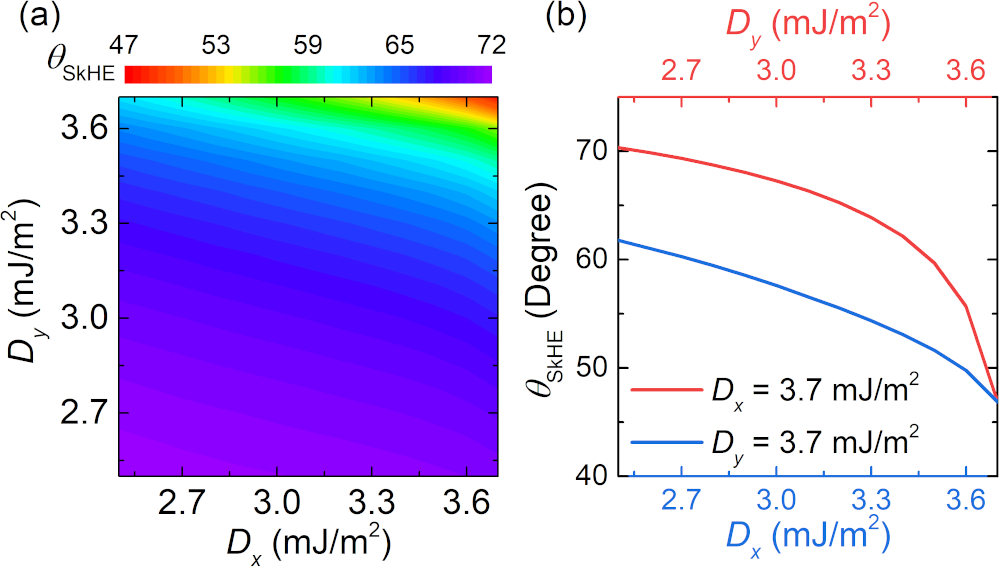}}
\caption{
(a) Skyrmion Hall angle as functions of $D_x$ and $D_y$.
(b) Skyrmion Hall angle as a function of $D_x$ when $D_y = 3.7$ mJ/m$^2$ and skyrmion Hall angle as a function of $D_y$ when $D_x = 3.7$ mJ/m$^2$.
}
\label{FIG3}
\end{figure}

We next investigate the motion of skyrmion driven by the spin-orbit torque.
Initially, the relaxed skyrmion is located at the center of the sample with a side length of $200$ nm. The spin current can be injected by utilizing the spin Hall effect in the heavy-metal substrate.
Figure~\ref{FIG3}(a) shows the skyrmion Hall angle $\theta_\text{SkHE}$ as functions of $D_x$ and $D_y$. It can be seen that $\theta_\text{SkHE}$ decreases with increasing $D_x$ and $D_y$.
In Fig.~\ref{FIG3}(b), for $D_x=3.7$ mJ/m$^2$, the skyrmion Hall angle decreases from $70.3^\circ$ to $46.9^\circ$ when $D_y$ increases from $2.5$ mJ/m$^2$ to $3.7$ mJ/m$^2$. The major axis of elliptical skyrmion is along the $y$-axis, as shown in Fig.~\ref{FIG2}(d). Similarly, for $D_y=3.7$ mJ/m$^2$, the skyrmion Hall angle decreases with increasing $D_y$. However, the major axis of  elliptical skyrmion is along the $x$-axis, as shown in Fig.~\ref{FIG2}(c).
It is noteworthy that the skyrmion Hall angle depends on the direction of the major axis of the elliptical skyrmion.
For example, the elliptical skyrmion has $a_\text{sk}=11.5$ nm and $b_\text{sk}=15.5$ nm when $D_x=3.7$ mJ/m$^2$ and $D_y=3.0$ mJ/m$^2$, of which $\theta_\text{SkHE}=67.3^\circ$. When $D_x=3.0$ mJ/m$^2$ and $D_y=3.7$ mJ/m$^2$, the elliptical skyrmion has $a_\text{sk}=15.5$ nm and $b_\text{sk}=11.5$ nm, of which $\theta_\text{SkHE}=57.6^\circ$.
Namely, the elliptical skyrmion with a major axis along the $x$-axis has a smaller skyrmion Hall angle compared with the one with the same area but a major axis along the $y$-axis.

\begin{figure}[t]
\centerline{\includegraphics[width=0.43\textwidth]{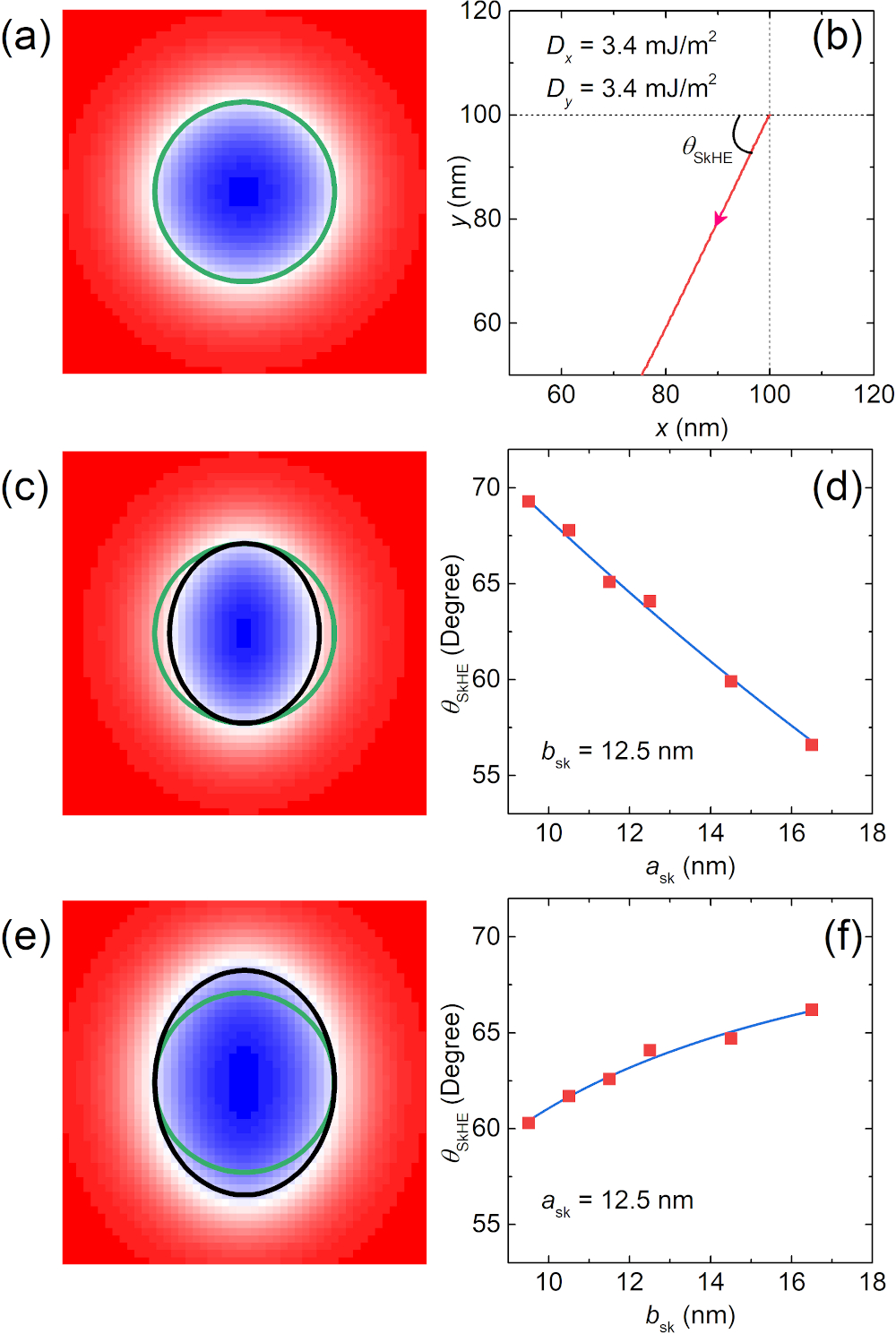}}
\caption{
(a) The top view and (b) trajectory of the skyrmion for the case of $D_x = D_y = 3.4$ mJ/m$^2$. The radius of the skyrmion equals $12.5$ nm. The skyrmion Hall angle $\theta_\text{SkHE}$ equals $64.1^\circ$.
(c) The top view of the skyrmion for the case of $D_x = 3.7$ mJ/m$^2$ and $D_y = 2.7$ mJ/m$^2$. $a_\text{sk}=9.5$ nm, $b_\text{sk}=12.5$ nm, and $\theta_\text{SkHE}=69.3^\circ$.
(d) Skyrmion Hall angle as a function of $a_\text{sk}$ when $b_\text{sk}=12.5$ nm.
(e) The top view of the skyrmion for the case of $D_x = 3.7$ mJ/m$^2$ and $D_y = 3.1$ mJ/m$^2$. $a_\text{sk}=12.5$ nm, $b_\text{sk}=16.5$ nm, and $\theta_\text{SkHE}=66.2^\circ$.
(f) Skyrmion Hall angle as a function of $b_\text{sk}$ when $a_\text{sk}=12.5$ nm.
The out-of-plane magnetization component is represented by the red ($+z$)-white ($0$)-blue ($-z$) color scale.
The green circle is the contour of the circle skyrmion ($m_z=0$) while the black oval is the contour of the elliptical skyrmion ($m_z=0$). The blue curves in (d) and (f) are the fitting results obtained by Eq.~\ref{eq:SHangle}.
}
\label{FIG4}
\end{figure}

Figure~\ref{FIG4}(a) shows the relaxed skyrmion for the case of $D_x = D_y = 3.4$ mJ/m$^2$. A circular skyrmion is obtained with a radius of $12.5$ nm. Figure~\ref{FIG4}(b) shows the straight trajectory of the circular skyrmion driven by the spin-orbit torque, of which the skyrmion Hall angle is equal to $64.1^\circ$ (see Supplementary Video 1).
Figure~\ref{FIG4}(c) shows the relaxed skyrmion for the case of $D_x = 3.7$ mJ/m$^2$ and $D_y = 2.7$ mJ/m$^2$. The anisotropic DMI leads to an elliptical skyrmion with $a_\text{sk}=9.5$ nm and $b_\text{sk}=12.5$ nm. The skyrmion Hall angle is $69.3^\circ$. Compared to the isotropic case, $\theta_\text{SkHE}$ is increased. The dependence of skyrmion Hall angle on $a_\text{sk}$ when $b_\text{sk}=12.5$ nm is shown in Fig.~\ref{FIG4}(d). It is found that $\theta_\text{SkHE}$ almost linearly decreases with increasing $a_\text{sk}$, which indicates that the skyrmion Hall angle decreases when the skyrmion is stretched in the $x$ direction.
When $a_\text{sk} = 16.5$ nm, the skyrmion Hall angle decreases to $56.6^\circ$ (see Supplementary Video 1).
Figure~\ref{FIG4}(e) shows the top view of the skyrmion when $D_x = 3.7$ mJ/m$^2$ and $D_y = 3.1$ mJ/m$^2$, where $a_\text{sk}=12.5$ nm and $b_\text{sk}=16.5$ nm. When the skyrmion is driven by the spin-orbit torque, the skyrmion Hall angle $\theta_\text{SkHE}=66.2^\circ$ (see Supplementary Video 1), which is larger than the one for the isotropic case ($64.1^\circ$). It can be seen from Fig.~\ref{FIG4}(f) that $\theta_\text{SkHE}$ increases when the skyrmion is stretched in the $y$ direction.
It should be mentioned that the same results can be obtained if the field-like torque is included (see Supplementary Information).

In order to understand the micromagnetic simulation results, we use the Thiele equation to describe the current-driven motion of skyrmion,~\cite{Thiele_PRL1973,Ado_PRB2017}
\begin{equation}
\boldsymbol{G}\times\boldsymbol{v}-\alpha\boldsymbol{\mathcal{D}}\cdot\boldsymbol{v}+\boldsymbol{p}\cdot\boldsymbol{\mathcal{B}}=\boldsymbol{0},
\label{eq:thiele}
\end{equation}
where $\boldsymbol{G}=(0,0,Q)$ with $Q=-\int d^{2}\mathbf{r}\cdot\boldsymbol{m}\cdot\left(\partial_{x}\boldsymbol{m}\times\partial_{y}\boldsymbol{m}\right)/4\pi$ and $\boldsymbol{m}=\boldsymbol{M}/M_\text{S}$ is the reduced magnetization. $\boldsymbol{v}$ is the velocity of the magnetic skyrmion. $\boldsymbol{\mathcal{D}}$ is the dissipative tensor.
The components are calculated by $\mathcal{D}_{\mu\nu}=\int d^{2}\mathbf{r}\left(\partial_{\mu}\boldsymbol{m}\cdot\partial_{\nu}\boldsymbol{m}\right)/4\pi$ where $\mu,\nu$ run over $x$ and $y$. For an elliptical skyrmion, $\mathcal{D}_{xx}\neq\mathcal{D}_{yy}$, $\mathcal{D}_{xy}=\mathcal{D}_{yx}=0$.
$\boldsymbol{\mathcal{B}}$ is the tensor relating to the driving force with $\mathcal{B}_{\mu\nu}=\frac{u}{a}\int d^{2}\mathbf{r}\cdot\left(\partial_{\mu}\boldsymbol{m}\times\boldsymbol{m}\right)_{\nu}/4\pi$. $u$ is the speed of the electrons and $a$ is the thickness of the sample. The value of $\mathcal{B}_{\mu\nu}$ can be evaluated for a given profile of a skyrmion.~\cite{Xia_PRAPPLIED2019}
Based on Eq.~\ref{eq:thiele}, we can obtain $\frac{v_y}{v_x}=\frac{Q}{\alpha \mathcal{D}_{yy}}$.

We further estimate the value of $\mathcal{D}_{yy}$. The magnetic skyrmion profile can be expressed as $\boldsymbol{m}(\boldsymbol{r})=\boldsymbol{m}(\theta,\phi)=(\sin\theta\cos\phi,\sin\theta\sin\phi,\cos\theta)$, where $\theta(\boldsymbol{r})$ linearly changes from $\pi$ at the center to $0$ at the edge.
Such an assumption holds true for nanoscale compact skyrmions.
$\phi(\boldsymbol{r})=Q_{\text{v}}\varphi+\eta$ and $\boldsymbol{r}=(x,y)=(a_\text{sk} r\cos\varphi, b_\text{sk} r\sin\varphi)$. $Q_{\text{v}}$ and $\eta$ are the vorticity and helicity of the magnetic skyrmion, respectively. In this work, the magnetic skyrmion with a skyrmion number of $Q=+1$ has a skyrmion vorticity of $Q_{\text{v}}=+1$ and helicity of $\eta=0$. Then, we can obtain that $\mathcal{D}_{yy}=\frac{\pi^2a_\text{sk}}{8b_\text{sk}}$. Therefore, the skyrmion Hall angle can be expressed as
\begin{equation}
\theta_\text{SkHE}=\arctan(\frac{v_y}{v_x})=\arctan(\frac{8b_\text{sk}}{\alpha\pi^2a_\text{sk}}).
\label{eq:SHangle}
\end{equation}
From Eq.~\ref{eq:SHangle}, it can be seen that the skyrmion Hall angle decreases with increasing $a_\text{sk}$ while increases with increasing $b_\text{sk}$. The fitting data with Eq.~\ref{eq:SHangle} are given in Fig.~\ref{FIG4}(d) and \ref{FIG4}(f) as blue curves. It can be seen that the analytical solutions are in line with the micromagnetic simulation results.


In conclusion, we have studied the current-driven motion of an elliptical skyrmion stabilized by the anisotropic DMI. It is found that the skyrmion Hall angle decreases with increasing $a_\text{sk}$, which indicates that the skyrmion Hall effect can be reduced when the major axis of the elliptical skyrmions is along the $x$-axis. However, the skyrmion Hall angle increases with increasing $b_\text{sk}$, which shows that the skyrmion Hall effect is enhanced when the major axis of the elliptical skyrmion is along the $y$-axis. Here, it should be mentioned that these results are valid when the spin current polarization is along the $+y$ direction, corresponding to the spin-orbit torque generated by the spin Hall effect.
In this case, the driving current should be applied along the long axis direction to minimize the skyrmion Hall angle. Nevertheless, for arbitrary spin current polarization or spin-orbit torque mechanism, it is always possible to find the best current direction to minimize the skyrmion Hall angle when the skyrmion is elliptical.
The minimization of skyrmion Hall angle is important for spintronic devices based on the in-line motion of skyrmions, such as the racetrack-type memory. The reason is that large skyrmion Hall angle may lead to the destruction of skyrmions at sample edges.
Also, it is worth mentioning that when the elliptical skyrmion is driven by the spin-transfer torque, similar results can be found (see Supplementary Information).

See Supplementary Information for top views of skyrmion for different values of $D_x$ and $D_y$, and the skyrmion Hall angle of the elliptical skyrmion driven by the spin transfer torque. The Supplementary Video 1 shows the current-driven motion of circular and elliptical skyrmions.

X.Z. was supported by the Presidential Postdoctoral Fellowship of The Chinese University of Hong Kong, Shenzhen (CUHKSZ).
M. E. acknowledges the support from the Grants-in-Aid for Scientific Research from JSPS KAKENHI (Grant Nos. JP18H03676, JP17K05490 and JP15H05854) and also the support from CREST, JST (Grant Nos. JPMJCR16F1 and JPMJCR1874).
X.L. acknowledges the support by the Grants-in-Aid for Scientific Research from JSPS KAKENHI (Grant Nos. JP17K19074, 26600041 and 22360122).
Y.Z. acknowledges the support by the President's Fund of CUHKSZ, Longgang Key Laboratory of Applied Spintronics, National Natural Science Foundation of China (Grant Nos. 11974298 and 61961136006), Shenzhen Fundamental Research Fund (Grant No. JCYJ20170410171958839), and Shenzhen Peacock Group Plan (Grant No. KQTD20180413181702403).




\end{document}